\documentclass[aps,prl,twocolumn]{revtex4-1}
\usepackage{graphicx}% Include figure files
\usepackage{amsmath}
\usepackage{dcolumn}% Align table columns on decimal point
\usepackage{bm}% bold math
\usepackage{float}

\begin{document}

\preprint{}

\title{Magneto-Optics of Exciton Rydberg States in a Monolayer Semiconductor}

\author{A. V. Stier$^{1*}$, N. P. Wilson$^{2*}$, K. A. Velizhanin$^{3}$, J. Kono$^{4}$, X. Xu$^2$, S. A. Crooker$^1$}

\affiliation{$^1$National High Magnetic Field Laboratory, Los Alamos, NM 87545, USA}
\affiliation{$^2$Department of Physics, University of Washington, Seattle, WA 98195, USA}
\affiliation{$^3$Theoretical Division, Los Alamos National Laboratory, Los Alamos, NM 87545, USA}
\affiliation{$^4$Departments of Electrical and Computer Engineering, Physics and Astronomy, and Materials Science and NanoEngineering, Rice University, Houston, TX 77005, USA}
\affiliation{$^*$these authors contributed equally}

\begin{abstract}
We report 65 tesla magneto-absorption spectroscopy of exciton Rydberg states in the archetypal monolayer semiconductor WSe$_2$. The strongly field-dependent and distinct energy shifts of the $2s$, $3s$, and $4s$ excited neutral excitons permits their unambiguous identification and allows for quantitative comparison with leading theoretical models. Both the sizes (via low-field diamagnetic shifts) and the energies of the $ns$ exciton states agree remarkably well with detailed numerical simulations using the non-hydrogenic screened Keldysh potential for 2D semiconductors. Moreover, at the highest magnetic fields the nearly-linear diamagnetic shifts of the weakly-bound $3s$ and $4s$ excitons provide a direct experimental measure of the exciton's reduced mass, $m_r = 0.20 \pm 0.01~m_0$.
\end{abstract}

\maketitle

The burgeoning interest in atomically-thin transition-metal dichalcogenide (TMD) semiconductors such as monolayer MoS$_2$ and WSe$_2$ derives in part from their direct optical bandgap and very strong light-matter coupling \cite{MakShan, XuReview}. In a pristine TMD monolayer, the fundamental optical excitation --the ground-state neutral ``A" exciton ($X^0$)-- can, remarkably, absorb $>$10\% of incoming light \cite{Li}. Moreover, in doped or highly excited monolayers distinct resonances due to charged excitons or multi-exciton states can develop in optical spectra \cite{Ross, ZWang1, Courtade, You, Falko, Hao}. The ability to spectrally resolve these and other features depends critically on material quality, which has markedly improved in recent years as techniques for synthesis, exfoliation, and surface passivation have steadily progressed.

The optical quality of exfoliated WS$_2$ and WSe$_2$ monolayers has recently improved to the point where signatures of the much weaker \emph{excited} Rydberg states of $X^0$ ($2s$, $2p$, $3s$, \emph{etc.}) have been reported based on various linear and nonlinear optical spectroscopies \cite{He, Ye, Chernikov, Zhu, Wang1, Hill, Stroucken}. Correct identification and quantitative measurements of excited excitons are of critical importance in this field, because they provide direct insight into the \emph{non}-hydrogenic attractive potential between electrons and holes that is believed to exist in 2D materials due to dielectric confinement and nonlocal screening \cite{Keldysh, Cudazzo, Berkelbach, Macdonald, Kyla}. This potential leads, for example, to an unconventionally-spaced Rydberg series of excited excitons and can generate an anomalous ordering of (\textit{s, p, d}) levels \cite{Ye}. Crucially, these excited states allow one to directly estimate the free-particle bandgap and binding energy of the $X^0$ ground state \cite{He, Ye, Chernikov, Zhu, Wang1, Hill}, both key material parameters that are otherwise difficult to measure in monolayer TMDs, and which are necessarily very sensitive to the surrounding dielectric environment \cite{Kyla, Latini, Stier_Nano, Raja}. Greatly desired, therefore, are incisive experimental tools for detailed studies of excited excitons in 2D semiconductors.

Historically, optical spectroscopy in high magnetic fields $B$ has provided an especially powerful way to identify and quantify excited excitons \cite{Miura, Knox, Hasegawa, Ritchie, Edelstein}, because each excited state shifts very differently with $B$. Crucially, these shifts can directly reveal fundamental parameters such as the exciton's mass, size, and spin -- essential information for benchmarking theoretical models.  For example, in the `weak-field limit' where the characteristic magnetic length $l_B = \sqrt{\hbar / eB}$ (=25.7/$\sqrt{B}$~nm) is much larger than an exciton's radius, optically-allowed excited excitons ($2s$, $3s$, ..., $ns$) can be uniquely identified by their different \textit{sizes}, which in turn are directly revealed via their quadratic diamagnetic shifts \cite{Miura, Knox, Stier_NatComm},
\begin{equation}
\Delta E_{\rm dia} = \frac{e^2}{8m_r} \langle r_\perp^2 \rangle B^2 = \sigma B^2 ~~ \textrm{    (if } l_B \gg r_{ns}).
\end{equation}
Here, $m_r$=$(m_e^{-1} + m_h^{-1})^{-1}$ is the exciton's reduced mass, $\sigma$ is the diamagnetic coefficient and $r_\perp$ is a radial coordinate perpendicular to $B$. The expectation value $\langle r_\perp^2 \rangle = \langle \psi_{ns} | r_\perp^2 | \psi_{ns} \rangle$ is calculated over the exciton's envelope wavefunction $\psi_{ns} (\textbf{r})$. The exciton's root-mean-square (rms) radius is therefore $r_{ns}$=$\sqrt{\langle r_\perp^2 \rangle}$=$\sqrt{8m_r \sigma}/e$. The key point is that excited states, being more loosely bound, are larger and therefore exhibit significantly larger diamagnetic shifts: \textit{e.g.}, in a 2D model with hydrogen-like Coulomb potential ($\sim$$1/r$), $\sigma_{2s}$ and $\sigma_{3s}$ are 39 and 275 times larger than $\sigma_{1s}$, respectively \cite{Ritchie}.

In the opposite `strong-field limit' where $l_B \ll r_{ns}$ and the spacing between Landau levels (LLs) exceeds typical binding energies, optically-allowed interband transitions effectively occur between LLs in the valence and conduction bands.  In conventional semiconductors, these transition energies therefore increase approximately linearly with $B$ as $ (N + \frac{1}{2})\hbar \omega_c^*$ (ignoring spin effects; $N=0, 1, 2...$), where $\hbar \omega_c^* = \hbar eB/m_r$ is the exciton's cyclotron energy. Importantly, this provides a direct experimental measure of $m_r$, independent of any model. Finally, in the intermediate regime where $l_B \sim r_{ns}$, a gradual crossover of $\Delta E_{\rm dia}$ from $B^2$ to $B$ dependence is expected \cite{Miura, Knox, Hasegawa, Ritchie, Edelstein}.  Magneto-optical studies of excited exciton states have a very successful history in III-V and II-VI semiconductors \cite{Miura, Knox}, and were employed 50 years ago to study bulk MoS$_2$ \cite{Evans}. To date, however, high-field studies of Rydberg excitons in the new family of monolayer TMDs has not been reported. 

Here we perform polarized magneto-optical spectroscopy to 65~T of monolayer WSe$_2$, an archetypal 2D semiconductor. The very different energy shifts of the $2s$, $3s$, and $4s$ excited states of $X^0$ are observed and studied for the first time.  This permits not only their unambiguous identification but also allows for direct quantitative comparison with leading theoretical models based on the non-hydrogenic screened Keldysh potential \cite{Berkelbach, Macdonald, Kyla}. A value of $m_r$ is experimentally obtained.

Figure 1a depicts the experiment. To achieve high optical quality, a single WSe$_2$ monolayer was sandwiched between 10~nm thick hexagonal boron nitride (hBN) slabs using a dry-transfer process and exfoliated materials.  The assembly was then affixed over the 3.5~$\mu$m diameter core of a single-mode optical fiber to ensure a rigid optical alignment. The fiber was mounted in the low-temperature (4~K) bore of a 65~T pulsed magnet.  Broadband white light from a Xe lamp was coupled through the structure via the single-mode fiber, and the transmitted light passed through a thin-film circular polarizer before being redirected back into a separate collection fiber. The collected light was dispersed in a 300~mm spectrometer and detected with a CCD detector.  Spectra were acquired every 2.3~ms throughout the magnet pulse, following \cite{Stier_Nano}.  Access to $\sigma^-$ or $\sigma^+$ circularly-polarized optical transitions (corresponding to transitions in the $K$ or $K'$ valley of WSe$_2$) was achieved by reversing $B$.

\begin{figure}[tbp]
\center
\includegraphics[width=0.45\textwidth]{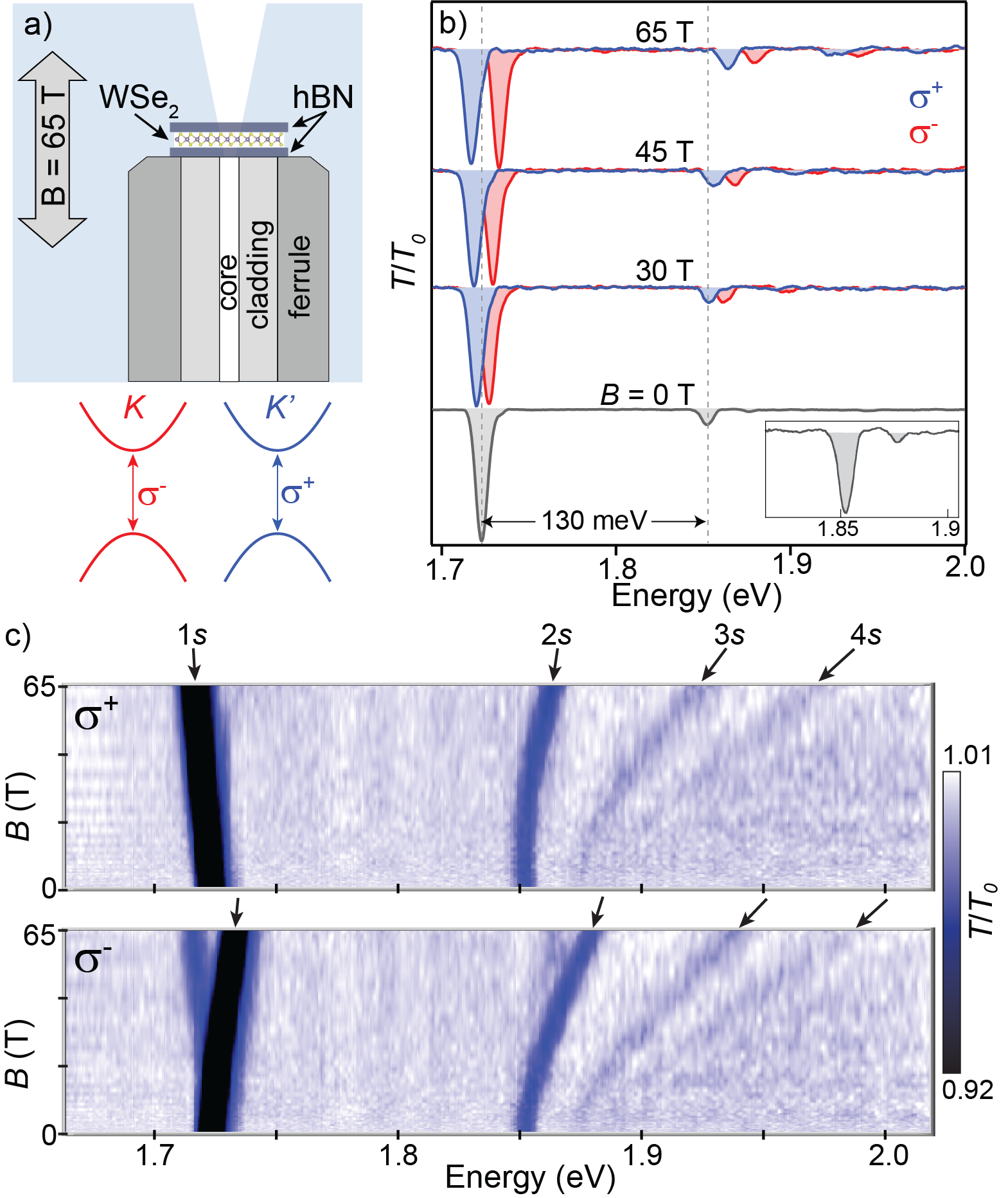}
\caption{(a) Experimental schematic: a WSe$_2$ monolayer, sandwiched between hBN slabs, is positioned over the 3.5~$\mu$m diameter core of a single-mode optical fiber. Circularly-polarized $\sigma^\pm$ transmission spectra are acquired to 65~T at low temperature (4~K). (b) Normalized transmission spectra, $T/T_0$, at selected magnetic fields $B$ from 0 to 65~T. The $1s$ ground  state of the neutral ``A" exciton $X^0$ appears at 1.723~eV. Its $2s$ excited state is also clearly visible at 1.853~eV (130~meV higher in energy); it exhibits a much larger diamagnetic blueshift in accord with its much larger spatial extent (vertical dashed lines indicate their zero-field energies). Inset: The $3s$ excited state is faintly visible even at $B$=0. (c) Intensity plots showing all the $\sigma^\pm$ spectra from 0-65~T. The large shifts of the weak $3s$ and $4s$ states of $X^0$ are readily apparent. (A small amount of $\sigma^+$ polarization leaks into the $\sigma^-$ spectra, especially for the strong $1s$ feature at large $B$.)}
\label{fig1}
\end{figure}

Figure 1b shows normalized transmission spectra ($T/T_0$) at 0, 30, 45, and 65~T. At $B$=0, the strong and narrow absorption line at 1.723~eV corresponds to the well-known  ground ($1s$) state of $X^0$. In addition, a weaker absorption also appears 130~meV higher in energy, at 1.853~eV. This feature has been observed in several studies of hBN-encapsulated WSe$_2$ monolayers \cite{Courtade, Scharf, Manca, Chow, Jin}, and has been ascribed either to the excited $2s$ state of $X^0$ \cite{Scharf, Manca}, or alternatively to a composite exciton-phonon resonance consisting of hBN and WSe$_2$ phonons coupled to the $X^0$ ground state \cite{Jin, Chow}. A central goal of this work is to elucidate the nature of this --and other-- higher energy states, based on their evolution in large $B$. 

As $B$ increases to 65~T, Fig.~1b shows that these absorption features split and shift. The Zeeman splitting and small diamagnetic shift of the $X^0$ ground state were observed previously in monolayer WSe$_2$ \cite{Stier_Nano}, albeit using different encapsulations. The similar splitting but much larger blueshift of the higher-energy absorption are clearly seen. Moreover, these spectra also reveal weak \textit{additional} features developing at even higher energy. To best visualize these changes, Fig.~1c shows an intensity map of all the $T/T_0$ spectra from 0-65~T.  A key result is that, in addition to the $X^0$ ground state and the smaller absorption at 1.853~eV, two additional absorption features are clearly discerned at higher energies, that blueshift even more rapidly with $B$. Based on their shifts and splittings (quantified in detail below), we can unambiguously associate these four features with the optically-allowed $1s$, $2s$, $3s$, and $4s$ Rydberg states of $X^0$.
\begin{figure*}[tbp]
\center
\includegraphics[width=.99\textwidth]{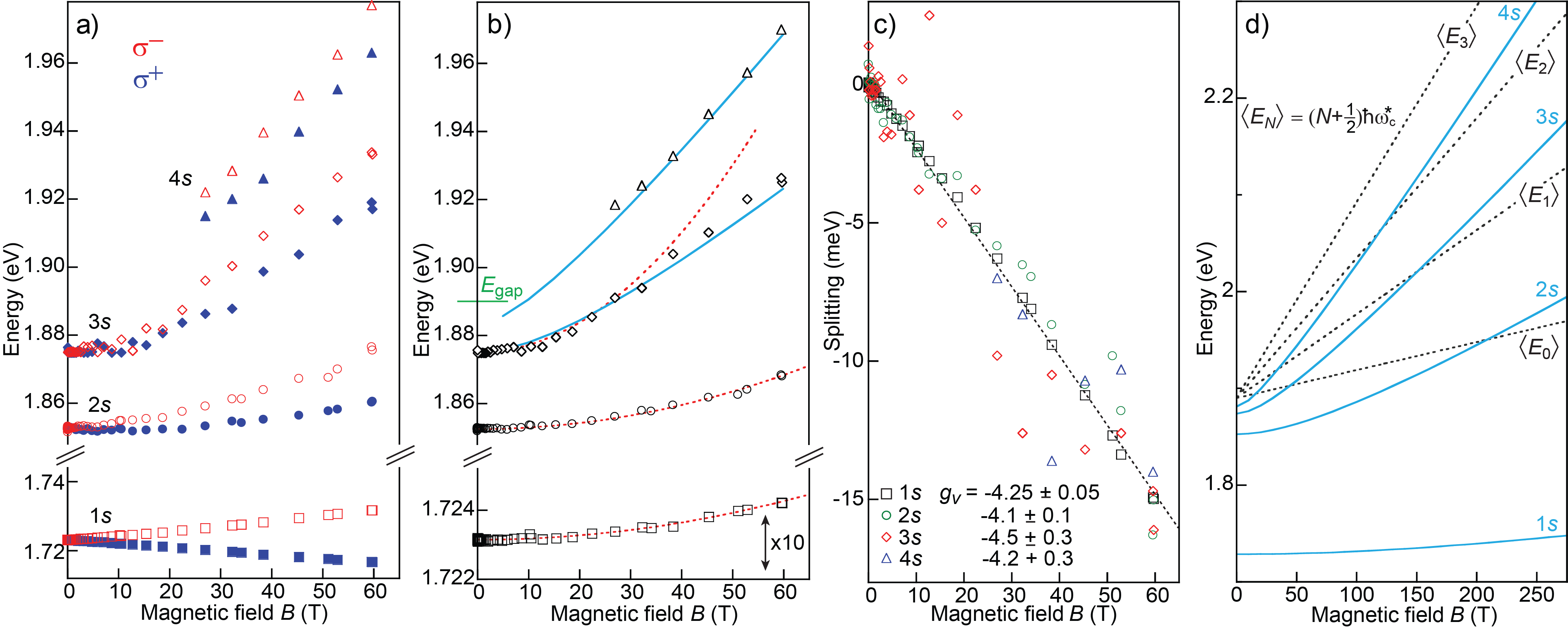}
\caption{(a) Measured $1s$, $2s$, $3s$, and $4s$ exciton energies versus $B$, for both $\sigma^+$ and $\sigma^-$ polarizations.  (b) The \textit{average} energy of the $\sigma^\pm$ transitions, for all the $ns$ states (note $10\times$ vertical scale for the $1s$ state). Dotted red lines show purely quadratic fits to $B^2$. The quadratic shifts of the $2s$ and $3s$ states are 15$\times$ and 71$\times$ larger than that of $1s$, quantitatively consistent with their larger radii computed from theory. The $3s$ and $4s$ states evolve toward a more linear shift at large $B$, which can be calculated numerically in this intermediate-field regime \cite{SM}. Blue lines show the numerically calculated $3s/4s$ energies using $m_r=0.2 m_0$. (c) The $\sigma^\pm$ energy difference reveals a similar valley Zeeman splitting for all $ns$ states.  The dotted straight line has slope $-245$~$\mu$eV/T ($g_v \approx -4.2$). (d) The blue lines show numerically calculated energies for all $ns$ Rydberg excitons to very large $B$. The straight dashed lines show $\langle E_N \rangle$=$(N + \frac{1}{2})\hbar \omega_c^*$, the valley-averaged energies of interband transitions between free electron and hole LLs \cite{SM}. At very large $B$, the slope of the $ns$ exciton shift approaches that of $\langle E_N \rangle$, where $N=n-1$.} \label{fig2}
\end{figure*}

Figure 2a quantifies these trends and shows the field-dependent $\sigma^\pm$ energies of these excitons, which follow $E(B) = E_0 + \Delta E_{\rm dia} \pm g_v \mu_B B$, where the last term describes the valley-dependent Zeeman splitting due to the exciton's magnetic moment \cite{Stier_NatComm}. The similar Zeeman splittings but very different diamagnetic shifts of the $ns$ excitons are readily apparent in Fig. 2a. Figure 2c shows the energy \textit{differences} between the $\sigma^\pm$ absorption peaks, revealing similar Zeeman splittings of $\sim$245~$\mu$eV/T, equivalent to a valley \textit{g}-factor $g_v \simeq -4.2$ for the $1s$ state of $X^0$ (in reasonable agreement with prior studies \cite{Srivastava, Wang, Stier_Nano, Mitioglu}), and also for the $ns$ excited states (measured here for the first time in a monolayer TMD). This concurrence is noteworthy because, as shown immediately below, the \textit{size} of these $ns$ excitons varies significantly by over an order of magnitude. Therefore their similar $g_v$ values point to a rather homogeneous distribution of orbital magnetism and Berry curvature in reciprocal space about the $K$ and $K'$ points of the Brillouin zone, in agreement with early theoretical studies of monolayer TMDs \cite{Cao}.

Most importantly, Fig.~2b shows the \textit{average} energy of the $\sigma^\pm$ absorption peaks for each $ns$ state, which reveals the diamagnetic shifts alone (independent of valley Zeeman effects).  The shift of the $1s$ exciton is small and purely quadratic ($\sigma_{1s}$=0.31$\pm$0.02~$\mu$eV/T$^2$, in line with recent studies of encapsulated WSe$_2$ \cite{Stier_Nano}), and directly reveals its small rms radius $r_{1s}$=1.7$\pm$0.1~nm via Eq.~1.  (Here we use $m_r$=0.20$m_0$, which is slightly larger than predicted by theory \cite{Kyla, Berkelbach}; however this value is consistent with modeling of $\sigma_{1s}$ \cite{SM} and as shown below is independently recommended by the high-field shifts of the $3s/4s$ states.)  In marked contrast to the $1s$ state, the quadratic shift of the $2s$ state is $\sim$15$\times$ larger ($\sigma_{2s} = 4.6 \pm 0.2$~$\mu$eV/T$^2$), confirming that the $2s$ exciton has a considerably larger radius $r_{2s} \simeq \sqrt{15}~r_{1s} \simeq$6.6~nm. Continuing, the $3s$ state exhibits an even more pronounced blueshift that follows $B^2$ up to 25~T. In this range, $\sigma_{3s} = 22 \pm 2$~$\mu$eV/T$^2$, which is $\sim$71$\times$ larger than $\sigma_{1s}$, indicating that $r_{3s} \simeq \sqrt{71}~r_{1s} = 14.3 \pm 1.5$~nm. These ratios ($\frac{\sigma_{2s}}{\sigma_{1s}}$=15 and $\frac{\sigma_{3s}}{\sigma_{1s}}$=71) are \textit{significantly} different than ratios expected from a hydrogenic exciton model in two dimensions (39 and 275, respectively \cite{Ritchie}), confirming that the effective Coulomb potential in real monolayer semiconductors deviates markedly from $1/r$.

Above 30~T, the $3s$ (and $4s$) energy shifts depart from $B^2$ and evolve towards a more linear dependence on $B$, indicating a crossover to the intermediate-field regime where $l_B \sim r_{3s} (r_{4s})$. As discussed below and at length in the Supporting Material \cite{SM}, their nearly-linear shifts at large $B$ can be used to experimentally determine $m_r$, values for which, to date, have been inferred primarily from density-functional theory \cite{Kyla, Berkelbach}. We note further that the oscillator strengths of the $3s$ and $4s$ excitons increase at large $B$ (see Fig. 1c), in accord with expectation \cite{Hasegawa}.

First, however, we show that the $15 \times$ and $71 \times$ larger diamagnetic shifts of the $2s$ and $3s$ excitons -- and also their zero-field energies of 130~meV and 152~meV above the $1s$ ground state -- agree remarkably well with straightforward modeling using the non-hydrogenic Keldysh potential that is believed to best describe electron-hole attraction in a monolayer material confined between dielectric slabs \cite{Keldysh, Cudazzo, Berkelbach, Macdonald, Kyla}:
\begin{equation}
V_K(r)=-\frac{e^2}{8 \varepsilon_0 r_0}\left[ H_0 \left(\frac{\kappa r}{r_0} \right) - Y_0 \left(\frac{\kappa r}{r_0}\right) \right].
\end{equation}
Here, $H_0$ and $Y_0$ are the Struve and Bessel functions of the second kind. The dielectric nature of the WSe$_2$ monolayer is characterized by its screening length $r_0 = 2 \pi \chi_{\rm 2D}$, where $\chi_{\rm 2D}$ is the 2D polarizability.  We use $r_0=4.5$~nm, consistent with theory \cite{Berkelbach, Kyla} and experimental work \cite{Stier_Nano}. The encapsulating hBN slabs are captured by $\kappa$, the average dielectric constant of the surrounding material: $\kappa$=$\frac{1}{2}(\epsilon_{\rm top} + \epsilon_{\rm bottom})$. We use high-frequency (infrared) dielectric constants, because the characteristic frequency at which a dielectric responds to an exciton is given roughly by its binding energy \cite{Bechstedt, Knox}, which is large in TMD monolayers.  Thus, we use $\kappa_{\rm hBN}$=4.5, based on infrared measurements \cite{Geick}. $V_K(r)$ scales as $1/\kappa r$ when $r \gg r_0$, but diverges only weakly as $\textrm{log}(r)$ when $r \ll r_0$, due to increased screening from the 2D material itself.  Eq. 2 is often used to model excitons in monolayer materials \cite{Chernikov, Berkelbach, Kyla, Macdonald}, and it approximates reasonably well the potentials derived from more advanced models \cite{Scharf, Latini}.

The black curve in Fig.~3a shows $V_K(r)$. Also shown are $\psi_{ns}(r)$, the $ns$ wavefunctions of $X^0$ calculated numerically via Schr\"odinger's equation using $m_r = 0.20 m_0$. The $1s$ ground state has a calculated binding energy of 161~meV, and radius $r_{1s} = \sqrt{\langle \psi_{\rm 1s} |r_\perp^2 | \psi_{\rm 1s} \rangle}$ = 1.67~nm which is very close to the value of 1.7~nm that was directly measured (in Fig. 2b) from $\sigma_{1s}$.  More importantly, we calculated $r_{2s}$=6.96~nm and $r_{3s}$=15.8~nm, which agree rather well (within 10\%) with the values of 6.6~nm and 14.3~nm that were directly measured from their diamagnetic shifts. Put another way, $\sigma_{2s}$ and $\sigma_{3s}$ in hBN-encapsulated monolayer WSe$_2$ are predicted to be 17$\times$ and 89$\times$ larger than $\sigma_{1s}$, in reasonable agreement with the 15$\times$ and 71$\times$ larger diamagnetic shifts that are experimentally measured, confirming their identity.

\begin{figure}[tbp]
\center
\includegraphics[width=.48\textwidth]{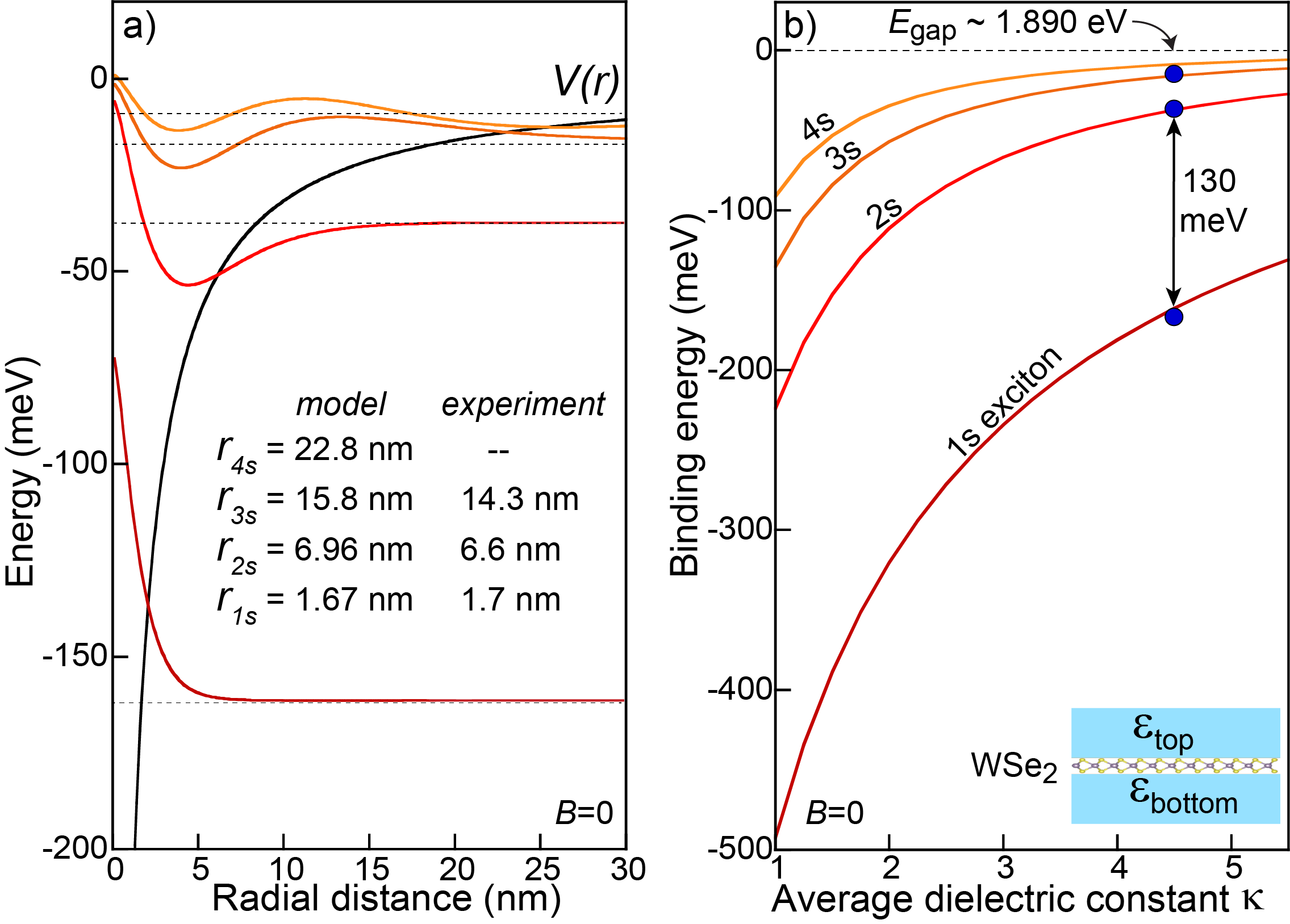}
\caption{(a) Plots of $\psi_{ns}(r)$, the $1s$, $2s$, $3s$, and $4s$ wavefunctions of $X^0$ in an hBN-encapsulated WSe$_2$ monolayer, computed using $V_K(r)$ (black line), using $r_0$=4.5~nm, $\kappa_{\rm hBN}$=4.5, and $m_r$=0.20$m_0$. The calculated rms exciton radii $r_{ns}$=$\sqrt{\langle \psi_{ns} | r_\perp^2 | \psi_{ns} \rangle}$ are close to experimental values. Crucially, we note that an rms radius is \textit{not} the same as a ``Bohr radius''; in the context of 3D (2D) hydrogenic models, rms radii are larger by a factor of $\sqrt{2}$ ($\sqrt{1.5}$) \cite{Stier_NatComm}. Thus the $1s$ rms exciton radius in \textit{bulk} WSe$_2$ was reported \cite{Beal1976} to be $\sqrt{2} \times 1.8$nm = 2.54~nm. (b) Calculated $ns$ exciton binding energies versus $\kappa$.  When $\kappa$=4.5, the calculated $1s$-$2s$ and $2s$-$3s$ separations are 124~meV and 21.3~meV, very close to the experimental values of 130~meV and 22~meV. Correlating the model with data (blue points) indicates a free-particle gap of $\sim$1.890~eV for hBN/WSe2/hBN.}
\label{fig3}
\end{figure}

This interpretation is further supported by Fig. 3b, which shows the calculated binding energies of the $ns$ excitons versus $\kappa$. The calculated $1s$-$2s$ and $2s$-$3s$ energy differences are 124~meV and 21.3~meV, respectively, when $\kappa$=$\kappa_{\rm hBN}$=4.5 \cite{Geick}. These values agree closely with the experimentally-measured separations of 130~meV and 22~meV, further confirming the nature of these Rydberg states and the applicability of $V_K(r)$ to monolayer TMDs. Overlapping the model with the measured exciton energies (blue points), we infer a free-particle bandgap $E_{\textrm{gap}} \approx 1.890$~eV for hBN/WSe$_2$/hBN.

Finally, we analyze the high-$B$ shifts of the $3s/4s$ excitons (Fig.~2b) to determine $m_r$, the reduced mass of $X^0$ -- a key material parameter that to date has not been directly measured. However, even at 65~T these excitons are only in the intermediate-field regime where their shifts are still evolving from quadratic to linear in $B$, and their energies lack simple analytic forms \cite{Miura, Knox, Hasegawa, Ritchie, Edelstein}. Nonetheless, the \emph{slopes} and \emph{separation} of the $3s/4s$ states at high $B$ provide upper and lower bounds on $m_r$, respectively \cite{SM}. The slope of the $4s$ shift, which should eventually \textit{increase} to $\frac{7}{2}\hbar \omega_c^* / B$ (see Fig.~2d), is $\sim$1.77~meV/T at 60~T, yielding an upper bound $m_r$$<$0.23$m_0$.  Conversely, the ratio $\delta/\hbar \omega_c^*$, where $\delta$ is the $3s$-$4s$ separation, should eventually \textit{decrease} to unity.  We measure $\delta$$\sim$34~meV at 60~T, giving a lower bound of $m_r$$>$0.16$m_0$. 

Tighter bounds on $m_r$ are obtained in this difficult intermediate-field regime by computing the exciton energies directly. However, analytical approximations have considered only hydrogen-like potentials \cite{Hasegawa, Ritchie}. Therefore, we numerically computed \cite{SM} the $B$-dependent exciton wavefunctions and energies using the relevant Hamiltonian for $s$-states in 2D semiconductors, $H = -(\hbar^2/2m_r)\nabla_r^2 + e^2 B^2 r^2/8m_r + V_K(r)$. In Fig.~2b we overlay these numerical results for the $3s$ and $4s$ states with the data.  A best fit is obtained using $m_r = 0.20 \pm 0.01 m_0$ (about 15\% larger than predicted by recent theory \cite{Berkelbach, Kyla}), thereby providing an internally-consistent experimental measure of $X^0$'s reduced mass in a monolayer TMD. 

Figure~2d shows the numerical results for all $ns$ states to very high $B$ ($>$250~T). Also plotted are the valley-averaged energies of the optically-allowed interband transitions between free-particle LLs, $\langle E_N \rangle = \frac{1}{2}(E_N^K + E_N^{K'})$=$(N + \frac{1}{2})\hbar \omega_c^*$ ($N$=0,1,2,...), which are analogous to inter-LL transitions in conventional semiconductors (for details, see the Supplemental Material \cite{SM}).  Only at extremely large $B$ ($\gg$100~T) are the $ns$ exciton shifts approximately parallel to those of $\langle E_N \rangle$ (where $N$=$n-1$), indicating the strong-field limit. Note that due to the exciton binding energy an offset always exists between the $ns$ exciton energy and the corresponding $\langle E_N \rangle$ energy. 

In summary, 65~T magneto-absorption spectroscopy was used to identify and quantify the optically-allowed $ns$ Rydberg states of neutral excitons in a monolayer semiconductor. The distinct shifts of the different $ns$ states allowed for direct quantitative comparison between experiment and theory. Both the sizes and energies of the $ns$ excitons are in good agreement with the screened Keldysh potential, and furthermore the nearly-linear energy shifts of the most weakly bound excitons provided an experimental measure of the exciton mass itself. Future studies using larger magnetic fields and/or higher-quality monolayers in which even higher Rydberg states are visible, can further improve these bounds in WSe$_2$ and other members of the monolayer TMD family.

We thank B. Urbaszek, X. Marie, and Wang Yao for helpful discussions. Work at the NHMFL was supported by NSF DMR-1157490, the State of Florida, and the US DOE. Work at U. Washington was supported by the Department of Energy, Basic Energy Sciences, Materials Sciences and Engineering Division (DE-SC0018171). K.A.V. was supported by the DOE BES EFRC program. J.K. was supported by AFOSR FA9550-14-1-0268.

\newpage
\onecolumngrid
\setcounter{figure}{0}
\setcounter{equation}{0}
\renewcommand{\thefigure}{S\arabic{figure}}
\subsection{Supplemental Material for Magneto-Optics of Exciton Rydberg States in a Monolayer Semiconductor}
\subsection{A. 2D Massive Dirac Hamiltonian}

The single-particle spectrum of electronic states in a monolayer transition-metal dichalcogenide (TMD) semiconductor is described by a two-dimensional (2D) massive Dirac Hamiltonian \cite{Xiao, Rose},
\begin{equation}
H=\frac{\Delta}{2}\sigma_z+v_F p_x\sigma_x+v_F p_y\sigma_y,\label{eq:mdh}
\end{equation}
where $\Delta$ is the bandgap and $v_F$ is the Fermi velocity. Pauli matrices are $\sigma_x$, $\sigma_y$, and $\sigma_z$. Momentum operators are defined as $p_{x}=-i\hbar\frac{\partial}{\partial x}$ and $p_{y}=-i\hbar\frac{\partial}{\partial y}$. Spin-orbit coupling is omitted in Eq. (\ref{eq:mdh}) as it simply amounts to shifting the bandgap depending on the spin-valley index \cite{Xiao}, and this work concerns only those bands involved in dipole-allowed optical transitions involving the uppermost valence band and the conduction band to which it is coupled (\textit{i.e.}, the lowest-energy ``A" exciton). This Hamiltonian is straightforwardly diagonalized by substituting the vector momentum operator with its eigenvalue, ${\bf p}\rightarrow\hbar{\bf k}$,  yielding a single conduction and a single valence band with energies
\begin{equation}
E_\pm=\pm \sqrt{\left(\frac{\Delta}{2}\right)^2+v_F^2 \hbar^2k^2} 
\end{equation}
and corresponding spinor wavefunctions $\Psi_\pm(\bf{k})$. For band-edge states where $v_F \hbar k\ll\Delta$, the parabolic expansion yields $E_\pm\approx\pm\left(\frac{\Delta}{2}+\frac{\hbar^2 k^2}{2m^*}\right)$ with electron/hole effective masses $m^*=\frac{\Delta}{2v_F^2}$. Corresponding spinor wavefunctions are $\Psi_+\approx (1,0)$ and $\Psi_-\approx(0,1)$. The Hamiltonian operator corresponding to these approximate band-edge energies and wavefunctions is the two-band effective mass Hamiltonian,
\begin{equation}
H=\begin{bmatrix}H_{c} & 0\\
0 & H_{v}
\end{bmatrix}=\begin{bmatrix}\frac{\Delta}{2}+\frac{p^{2}}{2m^{*}} & 0\\
0 & -\left(\frac{\Delta}{2}+\frac{p^{2}}{2m^{*}}\right)
\end{bmatrix},
\end{equation}
describing two non-interacting parabolic bands. Accordingly, if interband transitions are not treated explicitly, an electron-hole pair can be described via a two-particle Hamiltonian $H=H_e+H_h$, where $H_e=\frac{\Delta}{2}+\frac{p_e^2}{2m^*}$ and $H_h=-H_v=\frac{\Delta}{2}+\frac{p_h^2}{2m^*}$. Interactions between the particles and with external fields can now be incorporated into this Hamiltonian. This simple parabolic form of the Hamiltonian remains accurate as long as interactions are weak so that the electron/hole kinetic energy is much smaller than $\Delta$. Specifically, in the case of attractive electron-hole interactions and non-vanishing external magnetic fields, the parabolic approximation remains accurate so long as the exciton binding energy and the magnetic (cyclotron) energy are small compared to the bandgap. For example, the explicit dependence of the spinor wavefunctions on the momentum $\hbar k$ in the non-parabolic case results in the Berry curvature and quantum geometric tensor effects. These effects were estimated to produce non-parabolic corrections to the exciton energies of the order of only $\sim$10 meV in TMDs \cite{Imamoglu,Xiao2015}.

\subsection{2D Excitons in a Magnetic Field}

According to the previous section, the low-energy non-interacting Hamiltonian for an electron-hole pair in a 2D semiconductor is
\begin{equation}
H=\frac{p_e^{2}}{2m_{e}}+\frac{p_h^{2}}{2m_{h}},
\end{equation}
where the constant energy $\Delta$ is omitted. Electron and hole effective masses are $m_{e}$ and $m_{h}$, respectively.  Technically, these masses are identical when obtained from the 2D massive Dirac Hamiltonian but we keep them different here for
generality. Interaction with a perpendicular magnetic field $B$ is introduced in the standard gauge-invariant manner by the Landau-Peierls substitution (in SI units),
\begin{equation}
{\bf p}_{e(h)}\rightarrow {\bf p}_{e(h)}\pm e{\bf A}_{e(h)},
\end{equation}
where vectors are shown in bold. The magnitude of the electron charge is denoted by $e$. 
The attractive electron-hole interaction within the 2D semiconductor is captured by the screened Keldysh potential energy $V_K(|{\bf r}_e-{\bf r}_h|)$ (Eq. 2 in the main text).  The Hamiltonian for an exciton in a magnetic field is then
\begin{equation}
H=\frac{\left({\bf p}_{e}+e{\bf A}_{e}\right)^{2}}{2m_{e}}+\frac{\left({\bf p}_{h}-e{\bf A}_{h}\right)^{2}}{2m_{h}}+V_K(r),
\end{equation}
where ${\bf A}_{e(h)}=B[\hat{{\bf z}}\times{\bf r}_{e(h)}]/2=\frac{B}{2}(x_{e(h)}\hat{{\bf y}}-y_{e(h)}\hat{{\bf x}})$
is the vector potential of a static perpendicular magnetic field in
the symmetric gauge. Unit vectors are marked by hats. The momentum operator is ${\bf p}_{e(h)}=-i\hbar\frac{\partial}{\partial{\bf r}}$. 

For the Schr\"{o}dinger equation $H\Psi=E\Psi$ we perform the gauge transformation $\Psi\rightarrow e^{if}\Psi$, which amounts to the substitution ${\bf p}_{e(h)}\rightarrow{\bf p}_{e(h)}+\hbar\nabla_{e(h)}f$ in the Hamiltonian. We choose $f=\frac{eB}{2\hbar}(x_{e}y_{h}-y_{e}x_{h})$ \cite{Walck-1998-9088}. With this gauge function the translational motion of the entire exciton decouples from its internal dynamics. The total momentum of the exciton is now conserved and we set it to zero, which results in the following Hamiltonian for the relative motion of the electron-hole pair,
\begin{equation}
H=\frac{\left({\bf p}+e{\bf A}\right)^{2}}{2m_{e}}+\frac{\left({\bf p}-e{\bf A}\right)^{2}}{2m_{h}}+V_K(r),
\end{equation}
where ${\bf p}=-i\hbar\frac{\partial}{\partial{\bf r}}$, ${\bf r}={\bf r}_{e}-{\bf r}_{h}$
and ${\bf A}_{e(h)}=B[\hat{{\bf z}}\times{\bf r}]/2$. This Hamiltonian
is axially symmetric so it can be rewritten exactly as 
\begin{equation}
H=-\frac{\hbar^{2}}{2m_r}\left(\partial_{r}^{2}+\frac{1}{r}\partial_{r}-\frac{m^{2}}{r^{2}}\right)+\frac{e^{2}B^{2}}{8m_r}r^{2}+V_K(r) +\frac{\hbar eB}{2}m\left(\frac{1}{m_{e}}-\frac{1}{m_{h}}\right),\label{eq:Hr}
\end{equation}
where $m_r=(m_{e}^{-1}+m_{h}^{-1})^{-1}$ is the exciton's reduced mass, and $m=0,\pm1,\pm2,...$ is the azimuthal quantum number. The obtained Hamiltonian is similar to those derived elsewhere \cite{Akimoto-1967-181,MacDonald-1986-8336,Edelstein-1989-7697,Walck-1998-9088,Miura-2008}. The Hamiltonian becomes especially simple for axially symmetric $s$-states ($m=0$),
\begin{equation}
H^{(s)}=-\frac{\hbar^{2}}{2m_r}\left(\partial_{r}^{2}+\frac{1}{r}\partial_{r}\right)+\frac{e^{2}B^{2}}{8m_r}r^{2}+V_K(r).\label{eq:HrS}
\end{equation}
Note that this Hamiltonian does not explicitly contain any spin- or valley-dependent Zeeman terms, and as such its solutions can be directly compared to the \textit{average} of the valley-dependent exciton energies that are experimentally measured using $\sigma^+$ and $\sigma^-$ optical polarizations. 

\subsection{Numerical Methods}

The Schr\"{o}dinger equation corresponding to Hamiltonian (\ref{eq:HrS})
has been solved numerically on a grid. More specifically, the second-order
differential equation has been split into two first-order equations
by introducing an unknown function
\begin{equation}
\varphi(r)=\partial_{r}\Psi(r).\label{eq:phiPsi}
\end{equation}
 Then, each differential equation was transformed into a system of
finite-difference linear equations. Both $\Psi(r)$ and $\varphi(r)$
are represented on the equidistant grid $\{r_{i}\}$ with $r_{1}=0$
and $r_{N}=r_{\rm max}$, and the finite-difference equations are 
evaluated between the grid points, so that Eq.~(\ref{eq:phiPsi})
transforms to $N-1$ linear equations
\begin{equation}
\frac{1}{2}\left[\varphi(r_{i+1})+\varphi(r_{i})\right]-\frac{\Psi(r_{i+1})-\Psi(r_{i})}{r_{i+1}-r_{i}}=0,\,i=1,2,..,N-1. \label{eq:eq1}
\end{equation}
The other differential equation transforms to $N-1$ linear equations
as
\begin{gather}
-\frac{\hbar^{2}}{2m_r}\left(\frac{\varphi(r_{i+1})-\varphi(r_{i})}{r_{i+1}-r_{i}}+\frac{1}{r_{i+1}+r_{i}}\left[\varphi(r_{i+1})+\varphi(r_{i})\right]\right)+\frac{e^{2}B^{2}}{8m_r}\left(\frac{r_{i+1}+r_{i}}{2}\right)^{2}\nonumber \\
+V_K\left(\frac{r_{i+1}+r_{i}}{2}\right)=E\left[\Psi(r_{i+1})+\Psi(r_{i})\right]/2,\,i=1,2,...,N-1. \label{eq:eq2}
\end{gather}
This scheme naturally avoids the original $r=0$ singularity
of the radial Hamiltonian, Eq.~(\ref{eq:HrS}), since no function is evaluated directly at $r = r_1 = 0$. Note that the above
finite-difference scheme yields $2N-2$ equations with $2N$ unknowns.
The two extra equations are boundary conditions: (i) vanishing wavefunction
at $r_{\rm max}$, \textit{i.e.}, $\Psi(r_{N})=0$, and (ii) vanishing derivative
at the origin, $\varphi(r_{1})=0$.

Eqs.~(\ref{eq:eq1}) and (\ref{eq:eq2}) together with the boundary conditions do not constitute a standard matrix eigenvalue problem since only some of the equations include an eigenenergy $E$. Instead, a generalized eigenvalue problem is formulated
\begin{equation}
\hat{A}x=E\hat{B}x,
\end{equation}
where the matrix $\hat{A}$ incorporates the left-hand sides of the finite-difference and boundary condition equations above, and $x$ is a vector of size $2N$ containing components of the radial wavefunction and its first derivative on the grid, $\left\{ \Psi(r_{i}),\varphi(r_{i})\right\} $. The matrix $\hat{B}$ is mostly zero except for its elements corresponding to the right-hand side of Eq.~(\ref{eq:eq2}). This generalized eigenvalue problem can then be solved by standard means in, \textit{e.g.}, Mathematica. The resulting energies are subject to convergence tests with respect to the step size of the grid (the smaller the better) and $r_{\rm max}$ (the larger the better).  Finally, we also verified that this matrix method agreed with solutions given by iterative numerical Runge-Kutta methods.

\subsection{Calculated Energies of Optically-Allowed $ns$ Rydberg Excitons in Monolayer TMDs}

Figure S1 shows the energies of the $1s$, $2s$, $3s$, and $4s$ excitons versus $B$, calculated numerically using the screened Keldysh potential $V_K(r)$ for the case of monolayer WSe$_2$ sandwiched between thick hBN slabs ($\kappa=\frac{1}{2}[\epsilon_{\rm top}+\epsilon_{\rm bottom}]=4.5$). For these and all subsequent calculations, we use the WSe$_2$ screening length $r_0 = 4.5$~nm and a reduced exciton mass $m_r=0.20 m_0$, where $m_0=9.11 \times 10^{-31}$~kg is the bare electron mass. The left panel shows results in the experimentally-measured range up to 65~T, while the right panel shows the same results extending out to much higher fields.  As discussed in the main text, these exciton energies increase as $B^2$ in the weak-field regime (this is the quadratic diamagnetic shift, which depends on the exciton's size), while at higher fields the exciton energies gradually transition to a more linear-in-$B$ dependence (especially noticeable for the most weakly-bound $3s$ and $4s$ states).
\begin{figure*}[h]
\center
\includegraphics[width=.9\textwidth]{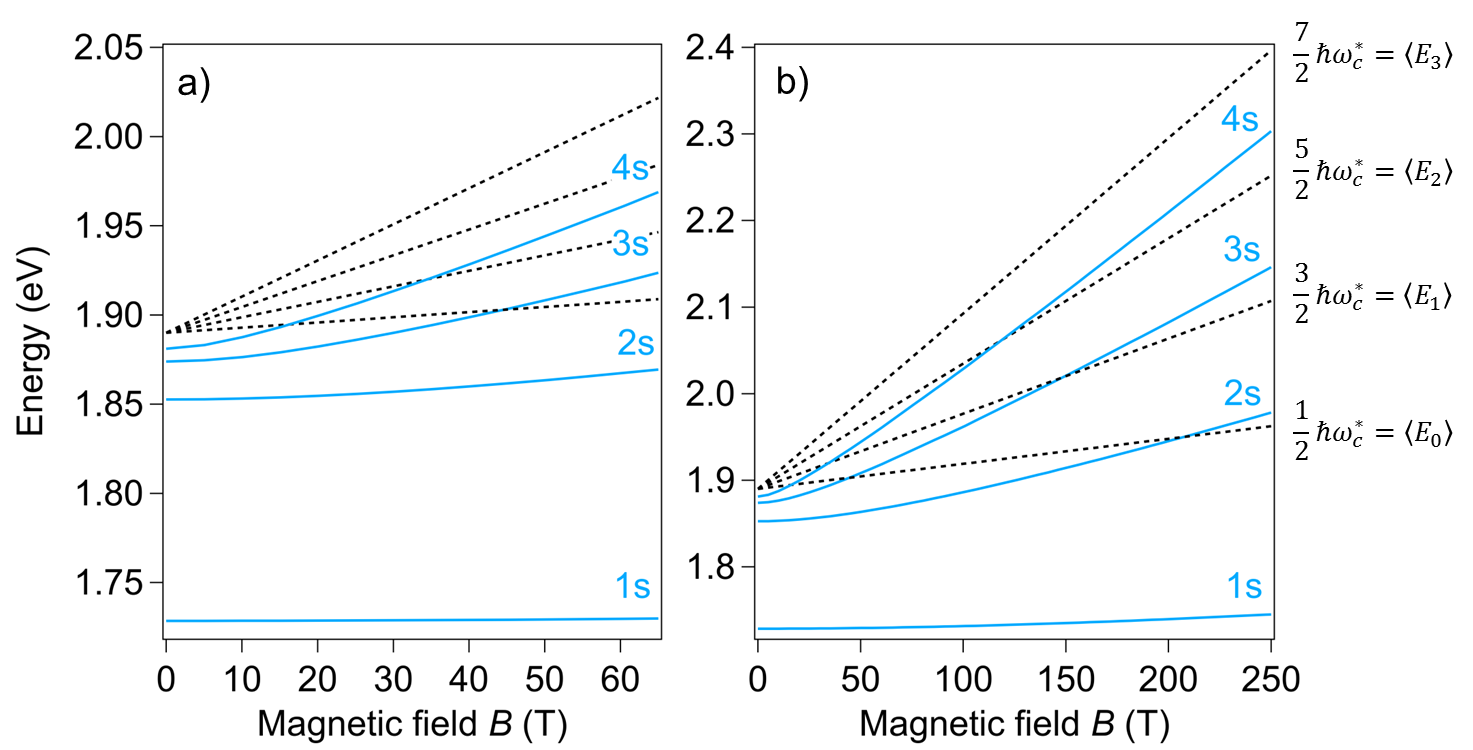}
\caption{The blue lines show the magnetic field dependence of the optically-allowed $1s$, $2s$, $3s$, and $4s$ exciton energies for hBN-encapsulated monolayer WSe$_2$ ($\kappa = 4.5$), calculated by numerically solving Schr\"{o}dinger's equation using the Hamiltonian (\ref{eq:HrS}) above and using $r_0 = 4.5$~nm and $m_r=0.2 m_0$. The left panel shows results in the experimentally-measured range up to 65~T; the right panel shows extended calculations up to 250~T. These calculations do not take into account any spin/valley-dependent Zeeman splitting; therefore these energies correspond to the \textit{average} of the experimentally-measured $\sigma^+$ and $\sigma^-$ optical transition energies (as shown, \textit{e.g.}, in Fig. 2b of the main text).  The straight black dashed lines correspond to half-odd-integer multiples of the exciton's cyclotron energy $\hbar \omega_c^* = \hbar e B/m_r$; \textit{i.e.}, $\frac{1}{2}\hbar \omega_c^*,~\frac{3}{2}\hbar \omega_c^*,~\frac{5}{2}\hbar \omega_c^*$, and $\frac{7}{2}\hbar \omega_c^*$. These four energies are equivalent to $\langle E_N \rangle$ ($N$=0, 1, 2, 3), the \textit{average} of the $K$ and $K'$ interband transition energies between free-particle Landau levels, for each of the four lowest optically-allowed inter-LL transitions in $K$ and $K'$ (see text).} \label{fig1}
\end{figure*}

Figure S1 also shows straight dashed black lines that correspond to half-odd-integer multiples of the exciton cyclotron energy $\hbar \omega_c^* = \hbar e B /m_r$.  As discussed immediately below, these energies are equivalent to the \textit{average} of the $K$ and $K'$ interband transition energies between the free-particle Landau levels in the conduction and valence bands. 

\subsection{Landau Levels in Monolayer TMDs}

Landau levels (LLs) form in the conduction and valence bands in the presence of an applied magnetic field $B$. The energies and dispersion of these Landau levels in 2D materials has been extensively discussed in the literature \cite{Li2, Chu, Wang_LL, Bieniek}; for the case of monolayer TMDs a particularly clear exposition can be found in Ref. \cite{Wang_LL}, which we adapt and extend here for completeness. 

\begin{figure*}[b]
\center
\includegraphics[width=.5\textwidth]{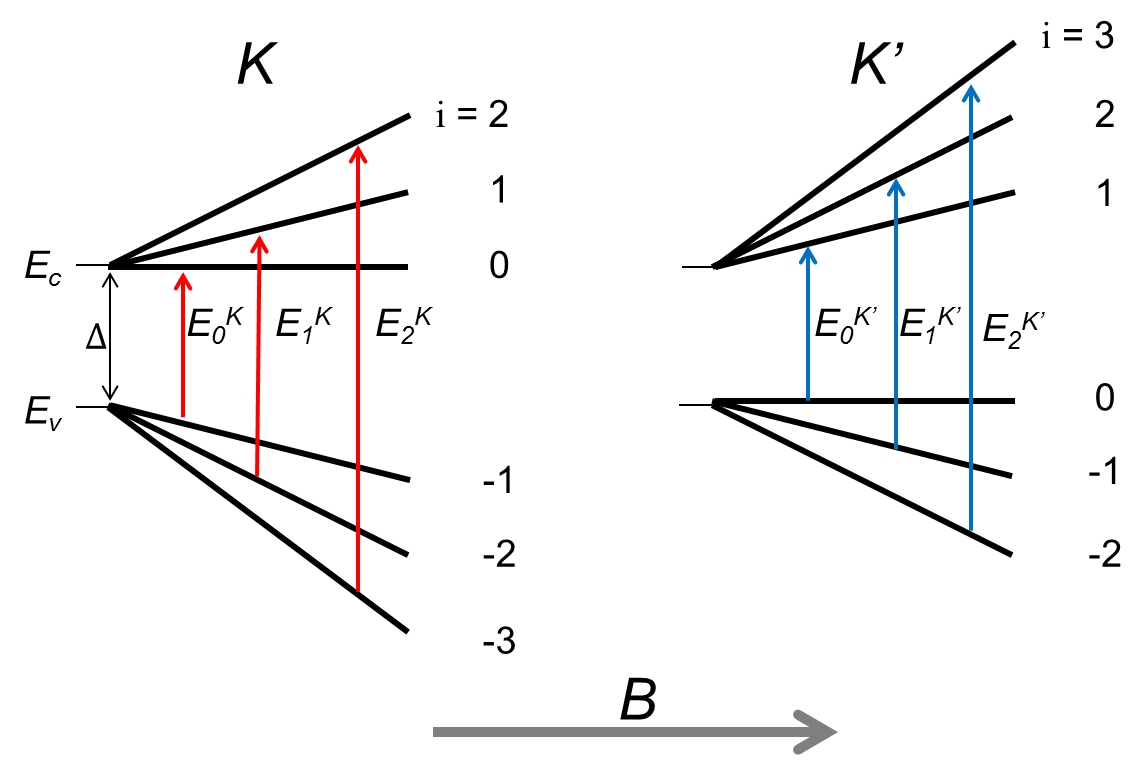}
\caption{Minimal diagram of the electron and hole Landau levels (LL) in the conduction and valence bands, in both the $K$ and $K'$ valleys of a monolayer TMD semiconductor. For clarity, the conduction and valence bands that do not participate in dipole-allowed optical transitions are not shown. Optically-allowed interband transitions are denoted by vertical arrows. Individual Landau levels are indexed by $i$, while the optically-allowed interband transition energies $E_N^K$ and $E_N^{K'}$ are labeled by the index $N$=0, 1, 2, ... from the lowest to highest energy. See text for details.} \label{fig2}
\end{figure*}

Figure S2 shows a minimal diagram of the free-particle LLs in the conduction and valence bands of a monolayer TMD semiconductor, along with vertical arrows denoting optically-allowed interband transitions. The zeroth LL in monolayer TMDs is pinned to the bottom of the conduction band in the $K$ valley, and is pinned to the top of the valence band in the $K'$ valley as depicted.  As discussed above, the parabolic-band approximation to the massive Dirac Hamiltonian holds as long as the characteristic cyclotron energies remain much smaller than the bandgap $\Delta$.  This limit holds very well (as shown later) for the magnetic fields and for the exciton Rydberg states considered in these studies.  Within this approximation, the LLs disperse linearly with $B$, analogous to conventional semiconductors.

In the conduction band of a monolayer TMD, the energy of the $i^{\rm th}$ electron LL is given by 
\begin{equation}
E_c + i \frac{\hbar e}{m_e}B \pm \mu_e B,\label{eq:Eci}
\end{equation}
where $E_c$ is the energy of the conduction band edge, the second term is the electron's cyclotron energy, and the third term is the Zeeman energy ($\mu_e$ is the magnetic moment of the conduction band and ``$\pm$" refers to the $K'$ and $K$ valleys, respectively). These electron LLs are therefore denoted by their index $i$, which runs from $i$=0, 1, 2, ... in the $K$ valley and $i$=1, 2, 3, ... in the $K'$ valley (typically these LLs are indexed by the letter ``$n$" in the literature, but we wish to avoid any potential confusion with our use of ``$ns$" to denote Rydberg excitons).

In the valence band, the energy of the $i^{\rm th}$ hole LL is given by
\begin{equation}
E_v + i \frac{\hbar e}{m_h}B \pm \mu_h B,\label{eq:Evi}
\end{equation}
where $E_v$ is the energy of the valence band edge, the second term is the hole's cyclotron energy, and the third term is the Zeeman energy ($\mu_h$ is the magnetic moment of the valence band and ``$\pm$" again refers to the $K'$ and $K$ valleys). These hole LLs are denoted by indices $i$ = -1, -2, -3, ... in the $K$ valley and $i$ = 0, -1, -2, ... in the $K'$ valley.  

Optically-allowed interband transitions in the $K$ valley, indicated by vertical arrows in Fig. S2, occur between LLs with $-(i+1)\leftrightarrow i$. We label these transitions from lowest to highest energy by the index $N$=0, 1, 2, ... . The energy of the $N^{\rm th}$ inter-LL optical transition in the $K$ valley is therefore
\begin{equation}
E_N^K = \Delta + \frac{\hbar e}{m_h}B + N\frac{\hbar e}{m_r}B - (\mu_e - \mu_h)B ~~~~~~ (N=0, 1, 2,...).
\end{equation}
where $\Delta = E_c - E_v$ is the free-particle bandgap. Note that this expression is slightly different from that in Ref. \cite{Wang_LL} because our index $N$ starts counting from 0 instead of 1. 

Similarly, optically-allowed interband transitions in the $K'$ valley occur between LLs with $-i \leftrightarrow i+1$ as depicted. Again labeling these transition from lowest to highest energy by $N$=0, 1, 2, ..., the energy of the $N^{\rm th}$ inter-LL optical transition in the $K'$ valley is 
\begin{equation}
E_N^{K'} = \Delta + \frac{\hbar e}{m_e}B + N\frac{\hbar e}{m_r}B + (\mu_e - \mu_h)B ~~~~~~ (N=0, 1, 2,...).
\end{equation}

The \textit{average} energy of the lowest interband transitions in the $K$ and $K'$ valleys is therefore
\begin{equation}
\langle E_0 \rangle = \frac{1}{2}(E_0^{K} + E_0^{K'}) = \frac{1}{2} \left( 2\Delta + \hbar e B \left( \frac{1}{m_e} + \frac{1}{m_h}\right) \right) = \Delta + \frac{1}{2}\hbar \omega_c^*.
\end{equation}
The average energy of the second-lowest interband transitions in the $K$ and $K'$ valleys is
\begin{equation}
\langle E_1 \rangle = \frac{1}{2}(E_1^{K} + E_1^{K'}) = \frac{1}{2} \left( 2\Delta + \hbar e B \left(\frac{1}{m_e} + \frac{1}{m_h}\right) + 2\frac{\hbar e}{m_r}B \right)  = \Delta + \frac{3}{2}\hbar \omega_c^*,
\end{equation}
and so on, where $\hbar \omega_c^* = \hbar e B/m_r$ is the exciton's characteristic cyclotron energy. Zeeman terms cancel out, and what remains is a linear dependence on $B$ with a slope equal to half-odd-integer multiples of $\hbar \omega_c^*$, similar to the case of conventional semiconductors.  In general, therefore,
\begin{equation}
\langle E_N \rangle = \frac{1}{2}(E_N^{K} + E_N^{K'}) = \Delta + \left(N+\frac{1}{2}\right)\hbar \omega_c^* ~~~~~~ (N=0, 1, 2,...).
\end{equation}

The straight dashed black lines in Fig. S1 depict the linear field dependencies of $\langle E_N \rangle$. With increasing $B$, the calculated energy shifts of the $ns$ excitons (blue lines) gradually become more linear and tend towards the slopes of $\langle E_N \rangle$, where $n= N+1$.  That is, the slope of the $4s$ exciton's energy shift tends asymptotically towards that of $\langle E_3 \rangle$, the slope of the $3s$ exciton's energy shift tends asymptotically toward that of $\langle E_2 \rangle$, and so on. This fact can be used, as discussed in the main text and as shown below, to provide an experimental bound on the exciton's reduced mass $m_r$. However, note that an offset will always exist between the energy of an $ns$ exciton and the energy of its corresponding average inter-LL transition energy $\langle E_N \rangle$ (where $n= N+1$); this difference reflects the electron-hole Coulomb binding energy \cite{Akimoto-1967-181, MacDonald-1986-8336, Edelstein-1989-7697, Miura-2008}.  Put another way, the excited $ns$ exciton states can be regarded as being composed of free electrons and holes in their respective LLs, with an energy modified by the Coulomb interaction.  

\subsection{Placing Bounds on the Exciton's Reduced Mass from High-Field Data}
Figure S3a shows how the slope of the $ns$ exciton's energy shift, measured at the largest experimentally-accessible magnetic field, can be used to provide an upper bound on $m_r$, a fundamental material parameter that has not, to date, been experimentally measured. Figure S3a shows the slopes of the $4s$ and $3s$ exciton's energy shift, which converge asymptotically (from \textit{below}) to the slopes of $\langle E_3 \rangle$ ($\equiv \frac{7}{2} \frac{\hbar e}{m_r}$) and $\langle E_2 \rangle$ ($\equiv \frac{5}{2} \frac{\hbar e}{m_r}$). At 65~T, inferring $m_r$ from the measured slope of the $4s$ state will overestimate $m_r$, thereby providing an upper experimental bound on $m_r$. As discussed in the main text, this procedure yielded the upper bound $m_r < 0.23 m_0$. The numerical calculations show that this bound is expected to be about 20\% larger than the correct value of $m_r$. 
\begin{figure*}[h]
\center
\includegraphics[width=.7\textwidth]{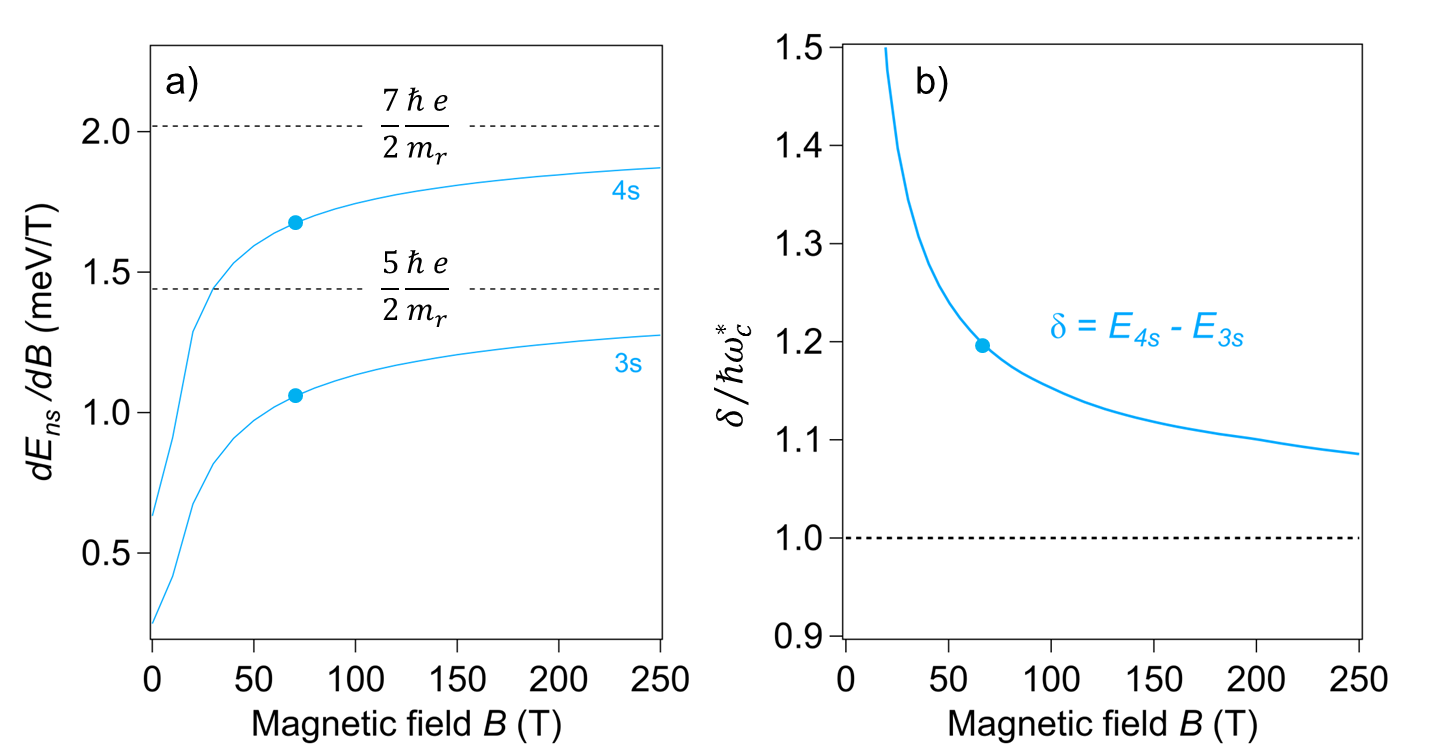}
\caption{Using the slopes and splitting between $ns$ exciton energies to place bounds on the exciton's reduced mass $m_r$. a) Calculated slope $dE_{ns}/dB$ of the $3s$ and $4s$ exciton energies versus $B$ (blue lines). These slopes asymptotically increase (from below) to the dashed black horizontal lines, which denote slopes of $\frac{5}{2}\hbar \omega_c^* / B$ and $\frac{7}{2}\hbar \omega_c^* /B$. The blue dots mark 65~T, the highest magnetic field used in these studies. The measured $4s$ slope at this point therefore yields a \textit{upper} bound on $m_r$ that is expected to deviate by about 20\% from the correct value. b) Calculated ratio of the $4s-3s$ energy difference ($\delta$) to the cyclotron energy ($\hbar \omega_c^*$). This ratio asymptotically decreases  to unity (from above).  Therefore the ratio measured at large $B$ yields a \textit{lower} bound on $m_r$ that is expected to deviate by about 20\% from the correct value.}  \label{fig1}
\end{figure*}

In contrast, Fig. S3b shows that the energy \textit{difference} between adjacent $ns$ exciton energies can be used to provide a lower experimental bound on $m_r$. Specifically, the ratio of this energy difference to $\hbar \omega_c^*$ is expected to decrease asymptotically to unity (from \textit{above}). Measuring this quantity at our highest experimentally-accessible field is therefore expected to underestimate $m_r$, thereby providing a lower experimental bound on $m_r$.  As discussed in the main text, this procedure yielded the lower bound $m_r > 0.16 m_0$.  The numerical calculations show that this bound is expected to be about 20\% smaller than the correct value of $m_r$. These bounds become narrower in higher $B$ and/or when higher $ns$ exciton states are visible in optical spectra.

Far more accurate, of course, is direct comparison of the measured $ns$ exciton energies with the full numerical calculation of the $ns$ exciton energies shown in Fig. S1.  As shown in the main text (Fig. 2b), very good agreement with the experimental data is achieved when $m_r = 0.20 m_0$.  

\subsection{Simulating the Influence of the Surrounding Dielectric Environment on $ns$ Rydberg Magneto-excitons}
The dielectric environment surrounding a TMD monolayer strongly influences exciton binding energies \cite{Stier_Nano} and therefore is also expected to influence the effect of an applied magnetic field $B$.  In Fig. S4 we show numerical calculations of the $ns$ exciton energies in monolayer WSe$_2$ for three common dielectric environments: a) freestanding in vacuum ($\kappa$=1), b) on a SiO$_2$ substrate ($\kappa$=1.5; we use the infrared value of silica, $\epsilon \sim 2$), and c) sandwiched between hBN slabs ($\kappa$=4.5; these latter calculations are the same as in Fig. S1). In all these plots, the zero-field energy of the $1s$ exciton was fixed to the same value of 1.723~eV, consistent with our measurements and in line with empirical observations that the $1s$ ground state exciton energy is largely independent of dielectric environment, due to the counteracting effects of reduced binding energy and correspondingly reduced free-particle gap \cite{Stier_Nano}.   
\begin{figure}[h]
\center
\includegraphics[width=.8\textwidth]{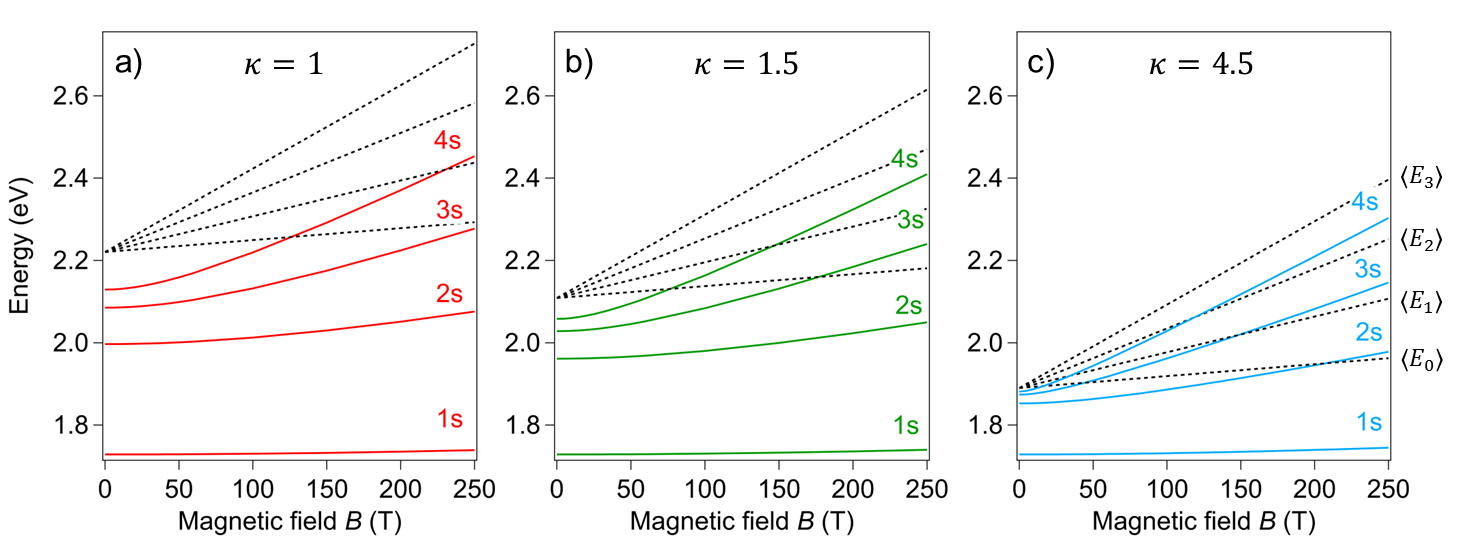}
\caption{Calculation of the $ns$ exciton energies, for monolayer WSe$_2$ in three different dielectric environments: a) freestanding in vacuum ($\kappa$=1), b) on a SiO$_2$ substrate ($\kappa$=1.5), and c) sandwiched between hBN slabs ($\kappa$=4.5). $\kappa=\frac{1}{2}(\epsilon_{\rm top}+\epsilon_{\rm bottom})$ is the average dielectric constant of the surrounding material.  We use $r_0 = 4.5$~nm and $m_r = 0.20 m_0$. To aid comparison and to correspond to experiments, we fix the $1s$ exciton energy at 1.723~eV at $B$=0.  As screening increases, exciton binding energies decrease (as does the free-particle gap), and at a given field the $ns$ exciton energies follow more closely their corresponding free-particle cyclotron energies $\langle E_N \rangle$ (where $n=N+1$). The dashed straight lines show $\langle E_N \rangle = (N+\frac{1}{2})\hbar \omega_c^*$.} \label{fig2}
\end{figure} 

The colored lines show the calculated $ns$ exciton energies, and as discussed above the straight black dashed lines show $\langle E_N \rangle = (N + \frac{1}{2})\hbar \omega_c^*$, which originate at $B$=0 at the free-particle gap. Increased dielectric screening reduces the exciton binding energies and also the free particle gap.  Increased screening also results in an earlier (lower $B$) transition from the weak-field regime to the intermediate-field regime where the exciton energies evolve away from a $B^2$ dependence and toward a more linear-in-$B$ dependence. Put differently, the strong-field regime is achieved at lower $B$ when excitons are more effectively screened by the surrounding dielectric, as expected. 

\subsection{Nonparabolicity Effects at High Field due to the 2D Massive Dirac Hamiltonian}

Landau levels within the 2D massive Dirac model do not disperse linearly with $B$. Rather, their energies are \cite{Rose}
\begin{equation}
E_i=\pm\sqrt{\left(\frac{\Delta}{2}\right)^2+i \left(\sqrt{2}\frac{\hbar v_F}{\l_0}\right)^2}, \label{eq:E_LLs}
\end{equation}
where $l_0=\sqrt{\hbar/eB}$ is the magnetic length. (For small $B$ where cyclotron energies are much less than the bandgap $\Delta$, it is easily verified that this expression yields the linearly-dispersing LLs computed above in Equations (\ref{eq:Eci}) and (\ref{eq:Evi}), where the particle mass is associated with $\frac{\Delta}{2 v_F^2}$, and ignoring Zeeman effects).  As discussed above, in the $K$ valley the $i$=0$^{\rm th}$ LL resides at the bottom of the conduction band and optically-allowed interband transitions occur between LLs with $-(i+1) \leftrightarrow i$. In the $K'$ valley the $i$=0$^{\rm th}$ LL resides at the top of the valence band and optically-allowed interband transitions occur between LLs with $-i \leftrightarrow i+1$. 

In Figure S5 the orange lines show the average of the $K$ and $K'$ interband transition energies, for each of the four lowest optically-allowed transitions, as calculated directly from the massive Dirac model using Equation (\ref{eq:E_LLs}). For comparison the straight black dashed lines show the corresponding interband LL transition energies computed in the parabolic approximation. As discussed in the first section, in the limit where the cyclotron energies are small compared to the bandgap, the parabolic approximation is valid and the transition energies disperse linearly as $\langle E_N \rangle = (N + \frac{1}{2})\hbar \omega_c^*$.  Nonparabolic corrections are very small in the experimentally accessible field range ($B<65$~T) and for the $ns$ excitons considered here.  At higher fields $>$100~T, however, such as those accessible using single-turn magnets or flux-compression magnet systems, nonparabolic effects are expected to become increasingly prominent, especially for highly excited (weakly bound) Rydberg excitons. 
  
\begin{figure}[h]
\center
\includegraphics[width=.5\textwidth]{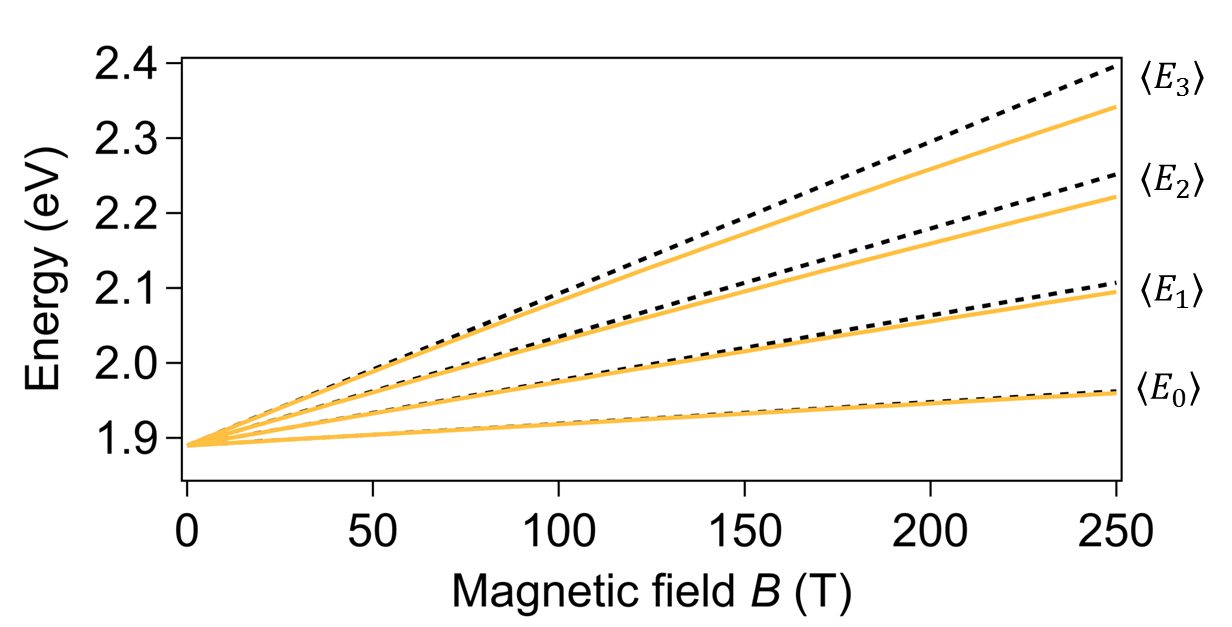}
\caption{Diagram comparing $\langle E_N \rangle = (N+\frac{1}{2})\hbar \omega_c^*$, the valley-averaged interband LL transition energies in the parabolic approximation (black dashed straight lines), with interband LL transition energies calculated via the massive Dirac Hamiltonian (orange solid lines). The band nonparabolicity predicted by the massive Dirac Hamiltonian is small in the experimentally accessible field range ($B<65$~T), but gradually increases at higher $B$ and is more pronounced for highly excited Rydberg excitons. Here we have used $m_e = m_h = 0.4 m_0$, and have ignored corrections due to trigonal warping \cite{Rose}.} \label{fig2}
\end{figure}

\subsection{A Self-Consistent Measure of the Exciton's Reduced Mass at Zero Field}

The measured quadratic diamagnetic shift of the $1s$ exciton, $\sigma_{1s}$, can be used to constrain estimates of the exciton's binding energy and reduced mass, as discussed previously \cite{Stier_Nano, Stier_NatComm}.  Specifically this is achieved by comparing $\sigma_{1s}$ with the diamagnetic shift that can be calculated by solving Schr\"{o}dinger's equation (utilizing the screened Keldysh potential $V_K(r)$ defined in the main text) to find $\psi_{1s}(r)$ and its rms radius $r_{1s}$. Following \cite{Stier_NatComm, Stier_Nano}, Fig. S6 shows a color map of the $1s$ exciton binding energy calculated for an hBN-encapsulated monolayer TMD, over a range of material screening lengths $r_0$ and reduced exciton masses $m_r$. Overlaid on the map is a \textit{contour of constant diamagnetic shift} that corresponds to the value of $\sigma_{1s} = 0.31$~$\mu$eV/T$^2$ that we experimentally measured for hBN-encapsulated monolayer WSe$_2$ (black line). This contour intercepts the expected WSe$_2$ screening length ($r_0 = 4.5$~nm) when $m_r=0.20 m_0$. This provides an internally consistent check on $m_r$ that agrees very well with the value of $m_r$ obtained from high-field measurements of the $3s$ and $4s$ excitons.  

\begin{figure}[h]
\center
\includegraphics[width=.4\textwidth]{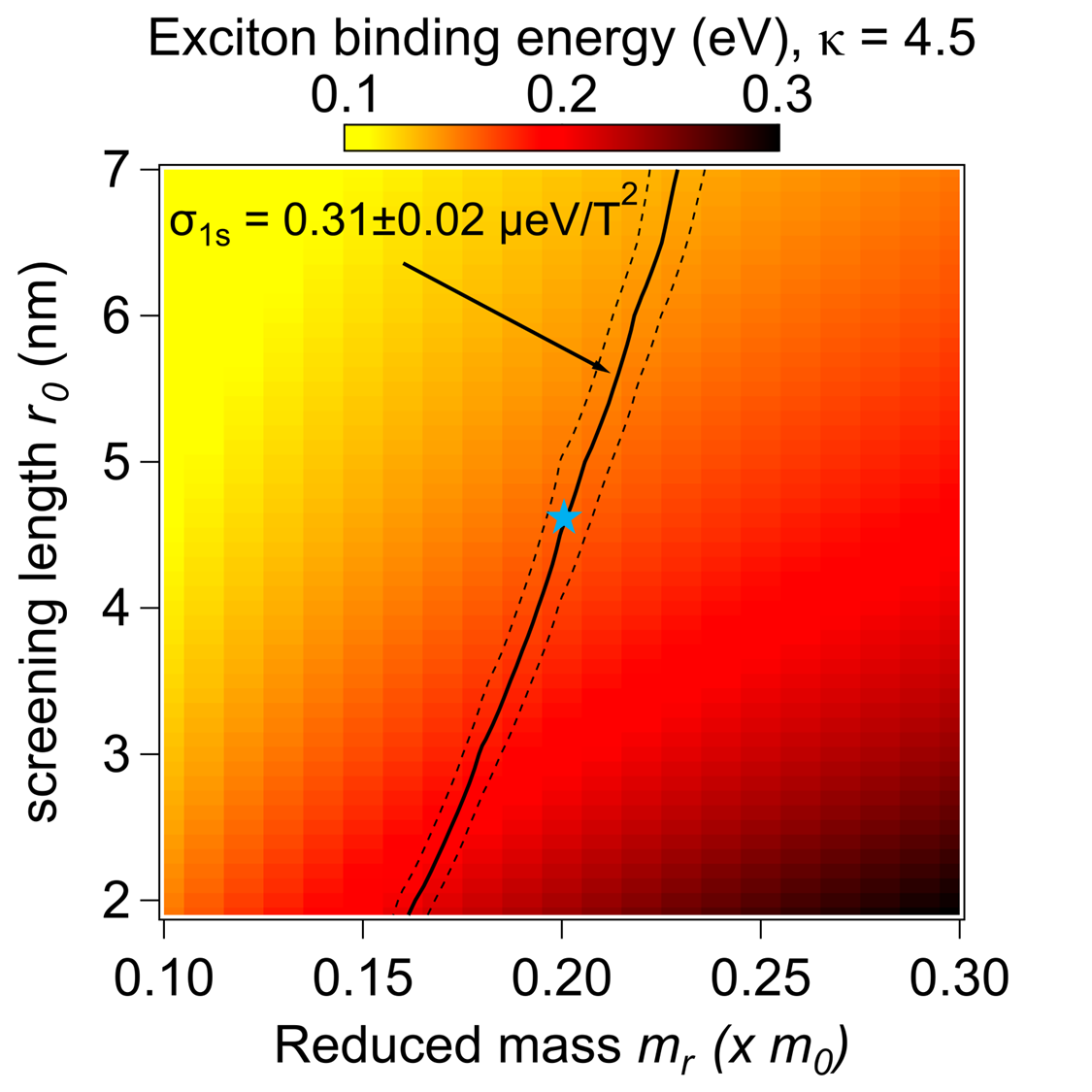}
\caption{Color surface plot of the $1s$ exciton binding energy in a monolayer TMD at $B=0$, calculated by solving Schr\"{o}dinger's equation using the screened Keldysh potential $V_K(r)$ with $\kappa = 4.5$ (corresponding to hBN encapsulation).  The calculations are performed over a range of possible reduced exciton masses $m_r$ and material screening lengths $r_0$. The black line denotes the contour of constant diamagnetic shift that corresponds to the experimentally-measured value of $\sigma_{1s} = 0.31$~$\mu$eV/T$^2$. Using the expected WSe$_2$ screening length $r_0 = 4.5$~nm, an exciton reduced mass of $m_r = 0.20 m_0$ is inferred.  This value is internally consistent with the value of $m_r$ obtained from high-field measurements of the $ns$ exciton shifts.} \label{fig3}
\end{figure}

\end{document}